\documentclass[preprint]{imsart}

\RequirePackage[OT1]{fontenc}
\RequirePackage{amsthm,amsmath}
\RequirePackage[numbers]{natbib}
\RequirePackage[colorlinks,citecolor=blue,urlcolor=blue]{hyperref}
\usepackage{amssymb}
\usepackage{amsthm}
\usepackage{graphicx}
\usepackage{algorithm}
\usepackage{enumerate}
\usepackage[left=3cm,right=3cm]{geometry}

\startlocaldefs
\numberwithin{equation}{section}
\theoremstyle{plain}
\newtheorem{lemma}{Lemma}[section]

\newtheorem{cor}{Corollary}[section]
\theoremstyle{remark}
\newtheorem{remark}{Remark}
\endlocaldefs

\begin{document}

\begin{frontmatter}
\title{Selection Bias Correction and Effect Size Estimation Under Dependence}
\runtitle{Effect Size Estimation}

\author{\fnms{Kean Ming}
  \snm{Tan}\corref{}\ead[label=e1]{keanming@uw.edu}} 
\address{Department of Biostatistics\\ University of Washington\\
Seattle, WA 98195, U.S.A.\\ \printead{e1}}
\author{\fnms{Noah}
  \snm{Simon}\corref{}\ead[label=e2]{nrsimon@uw.edu}} 
\address{Department of Biostatistics\\ University of Washington\\
Seattle, WA 98195, U.S.A.\\ \printead{e2}}
\and 
\author{\fnms{Daniela} \snm{Witten}\ead[label=e3]{dwitten@uw.edu}}
\address{Department of Statistics and Biostatistics \\ University of Washington\\
  Seattle, WA 98195, U.S.A.\\ \printead{e3}}

\runauthor{Tan et. al.}

\begin{abstract}
We consider large-scale studies in which it is of interest to estimate a very large number of effect sizes.  For instance, this setting arises in the analysis of gene expression or DNA sequencing data.     However, naive estimates of the effect sizes suffer from selection bias, i.e., some of the largest naive estimates are large due to chance alone.  Many authors have proposed methods to reduce the effects of selection bias under the assumption that the naive estimates of the effect sizes are independent.  Unfortunately, when the effect size estimates are dependent, as is often the case, these existing techniques can have very poor performance.  We propose an estimator that adjusts for selection bias under a recently-proposed frequentist framework, without the independence assumption.  We study some properties of the proposed estimator, and  illustrate that it outperforms past proposals in simulation studies and on two gene expression data sets.
\end{abstract}

\end{frontmatter}

\section{Introduction}
\label{Sec:Introduction}

In many applications, it is of interest to test a large number of hypotheses simultaneously.  For instance, in the context of a study in which tens of thousands of gene expression levels are measured in two groups of patients, one goal is to identify a subset of genes for which we can reject the null hypothesis of no mean difference between the groups \citep{dudoit2003multiple}.  A more ambitious goal is to estimate the effect sizes for the non-null features   \citep{montazeri2010shrinkage}.  Investigators often perform follow-up studies on the features with the largest estimated effect sizes.  Accurate effect size estimates are needed in order to ensure that follow-up studies are sufficiently powered, and to ensure that follow-up studies are performed only on hypotheses for which the effect size is large enough to be of practical interest.

Let $\delta_j$ be the true effect size for the $j$th hypothesis and let $\hat{\delta}_j$ be an  estimate of $\delta_j$.  Throughout the manuscript, we call the $\hat{\delta}_j$'s  the \emph{unadjusted estimates}.  For instance, in the context of differential expression testing, the unadjusted estimate $\hat{\delta}_j$ might be a two-sample $t$-statistic, and $\delta_j$ a standardized mean difference.

In practice, especially when the number of tested hypotheses is large, a given $|\hat{\delta}_j|$ might be large not only because its corresponding effect size $|\delta_j|$ is large, but also by chance.    Hence, the largest $|\hat{\delta}_j|$'s tend to overestimate the corresponding absolute effect sizes.
This is known as \emph{selection bias}.

Recently, several authors have proposed to correct for the effects of  selection bias  under the assumption that the unadjusted estimates are independent (among others, \citealp{efron2011tweedie,ferguson2013empirical, noah2013bias,sun2011br,xu2011bayesian,zhong2010correcting}).  This assumption is unrealistic in many real world settings, in which there may be dependencies among the unadjusted estimates. Gene expression data, in which features are known to be highly correlated (among others, \citealp{efron2007correlation,efron2010correlated,owen2005variance}), provides one such example.

In this paper, we propose two approaches for correcting selection bias in the case when the unadjusted estimates are dependent.  We build upon a recent proposal by \citet{noah2013bias}.  We account for dependencies among the unadjusted estimates using: (1) a parametric bootstrap, or (2) a nonparametric bootstrap.   We illustrate in simulation studies that failure to consider the dependencies among the unadjusted estimates can give inaccurate effect size estimates, and that our proposed approaches can overcome this problem.

One of the major strengths of our proposal is that we largely avoid the parametric assumptions about the unadjusted estimates made by existing proposals. Consequently, our proposal can be applied to a broad class of problems and statistics.  In particular, our proposal does not require normality of the data or of the unadjusted estimates.  However, in order to make a fair comparison to existing proposals, which \emph{do} require normality, we consider one-sample and two-sample $t$-statistics in our simulation studies and real data applications throughout the manuscript.

This paper is organized as follows.  In Section~\ref{Sec:Previous}, we review some previous work.  We describe our proposal in Section~\ref{Sec:Proposal}.  We study some properties of our proposed estimator in Section~\ref{Sec:Bias Correction Theory}.  In Section~\ref{Sec:Onesample}, we consider estimating the effect sizes of one-sample and two-sample $t$-statistics.  We illustrate our proposal on two gene expression data sets in Section~\ref{Sec:Realdata}.  We close with a discussion in Section~\ref{Sec:Discussion}.

\section{Previous Work}
\label{Sec:Previous}
In this section, we review the proposals of \citet{efron2011tweedie} and \citet{noah2013bias}.  Both approaches assume that the unadjusted estimates are independent.  

\subsection{Empirical Bayes Approach \protect \cite{efron2011tweedie}}
\label{subSec:Empirical}
\citet{efron2011tweedie} proposed to correct for selection bias using an empirical Bayes approach.    Let $g(\cdot)$ be some prior distribution and suppose that 
\begin{equation}
\label{Eq:EmpiricalBayes}
\delta_j \sim g(\cdot) \qquad \mathrm{and} \qquad \hat{\delta}_j|\delta_j \stackrel{\text{ind}}{\sim} f_{\delta_j}(\hat{\delta}_j) \qquad \text{for } j=1,\ldots,p,
\end{equation}
where $p$ is the number of effect sizes being estimated. The posterior expectation $E[\delta_j| \hat{\delta}_j]$ is immune to selection bias in the sense discussed in the Introduction \citep{dawid1994selection,senn2008note}.  If $f_{\delta_j}(\cdot)$ is the density of some exponential family distribution, then $E[\delta_j|\hat{\delta}_j]$ takes a simple form \citep{efron2011tweedie,robbins1985empirical}.  

In particular, under the additional assumption that $\hat{\delta}_j |\delta_j \stackrel{\text{ind}}{\sim} N(\delta_j,1)$, the posterior expectation is 
\begin{equation}
\label{Eq:posterior}
E[\delta_j | \hat{\delta}_j ] =\hat{\delta}_j+  \frac{d}{d\hat{\delta}_j}{\log f(\hat{\delta}_j)},
\end{equation}
where $f(\hat{\delta}_j) = \int g(\delta_j) f_{\delta_j} (\hat{\delta}_j)  d\delta_j$ is the marginal density of $\hat{\delta}_j$.
\citet{efron2011tweedie} suggested using the estimator 
\begin{equation}
\label{Eq:Empirical}
\hat{\delta}_j + \frac{d}{d\hat{\delta}_j} \log \hat{f} (\hat{\delta}_j),
\end{equation}
where $\log \hat{f} (\hat{\delta}_j)$ is an empirical estimate of the log marginal density of $\hat{\delta}_j$ obtained using Lindsey's method \citep{efron2008microarrays,efron2010large}.  We refer to this estimator as \verb=tweedie=\footnote{\citet{robbins1985empirical} credits personal correspondence with Maurice Kenneth Tweedie for the simple Bayesian estimation formula in Equation~\ref{Eq:posterior}.}.    More recently, \citet{wager2013geometric} proposed a more efficient estimate of $\log {f} (\hat{\delta}_j)$ using a non-linear projection approach, which we refer to as \verb=nlpden=.

\subsection{Frequentist Approach \protect \cite{noah2013bias}}
\label{subSec:noah}
We now review the proposal of \citet{noah2013bias}.  We start by defining some notation that will be used throughout the manuscript.  Let $\hat{\delta}_{(k)}$ denote the $k$th order statistic of the unadjusted estimates, $\hat{\delta}_{(1)} < \hat{\delta}_{(2)}< \cdots < \hat{\delta}_{(p)}$, assuming that there are no ties among the unadjusted estimates.  Define $j(k)$ as the index corresponding to the $k$th order statistic, i.e., $\hat{\delta}_{j(k)} = \hat{\delta}_{(k)}$.  For instance, if the fifth unadjusted estimate is the largest, then $j(p)=5$.

\citet{noah2013bias} defined the \emph{frequentist selection bias} of the $k$th order statistic $\hat{\delta}_{(k)}$  as 
\begin{equation}
\label{Eq:Bias}
\beta_k = E[\hat{\delta}_{(k)}-\delta_{j(k)}],
\end{equation}
where $j(k)$ is a random index since the ordering of $\hat{\delta}_j$'s is random.  Intuitively, $\beta_k$ quantifies the difference between the $k$th smallest unadjusted estimate and its corresponding true effect size; we expect this to be large when $k$ is large and small (negative) when $k$ is small.  
If the biases $\beta_1,\ldots,\beta_p$ were known, we would estimate  the effect size corresponding to $\hat{\delta}_{(k)}$ using the estimator 
\begin{equation}
\label{Eq:Frequentist}
\bar{\delta}_{j(k)}= \hat{\delta}_{(k)} - \beta_k.
\end{equation} 
We call this the \verb=oracle= estimator throughout the text.  Of course, the biases are unknown in practice.  

\citet{noah2013bias} provided a parametric bootstrap approach to estimate $\beta_k$ in Equation~\ref{Eq:Bias} under the assumption that $\hat{\delta}_j \stackrel{\text{ind}}{\sim} N(\delta_j,1)$.  They also hint about a generalization of this approach for non-normal and dependent unadjusted estimates.  We make this explicit in Section~\ref{subSec:parametric}.  
Given an estimate $\hat{\beta}_k$ of ${\beta}_k$ in Equation~\ref{Eq:Bias}, the proposed estimator of \citet{noah2013bias} takes the form
\begin{equation}
\label{Eq:ourestimator}
\tilde{\delta}_{j(k)} = \hat{\delta}_{(k)} - \hat{\beta}_k.
\end{equation}

\subsection{Dependent Unadjusted Estimates}
In practice, there are often dependencies among the unadjusted estimates.  We will show in later sections that failure to model these dependencies can lead to inaccurate estimates of the effect sizes.  

Existing empirical Bayes approaches for correcting effect size estimates for selection bias assume that the unadjusted estimates are independent \citep{efron2011tweedie}.  Assuming dependencies among the unadjusted estimates poses problems, since the posterior expectation cannot be simplified  as in Equation~\ref{Eq:posterior}, even under the normality assumption.  In principle, one could modify these existing approaches to accommodate dependence.  But it is not immediately obvious whether this is theoretically or computationally tractable.

In the next section, we propose a frequentist approach for correcting selection bias that accounts for  dependencies among the unadjusted estimates. This proposal extends ideas from  \citet{noah2013bias}. 

\section{Frequentist Selection Bias Under Dependence}
\label{Sec:Proposal}
We provide a simple approach to correct for selection bias in the presence of dependencies among the unadjusted estimates, using the framework of \citet{noah2013bias}.  As in Equation~\ref{Eq:ourestimator}, we consider the estimator $\tilde{\delta}_{j(k)} = \hat{\delta}_{(k)} - \hat{{\beta}}_k$.
The crux of our approach involves the technique used to calculate $\hat{\beta}_k$, as described  in Algorithm~\ref{Alg:process}.  Briefly, we estimate $\beta_k$ by computing the empirical selection bias over a large number of bootstrapped data sets for which the true effect size is known.

\begin{figure}[htp]
\begin{center}
\includegraphics[scale=0.475]{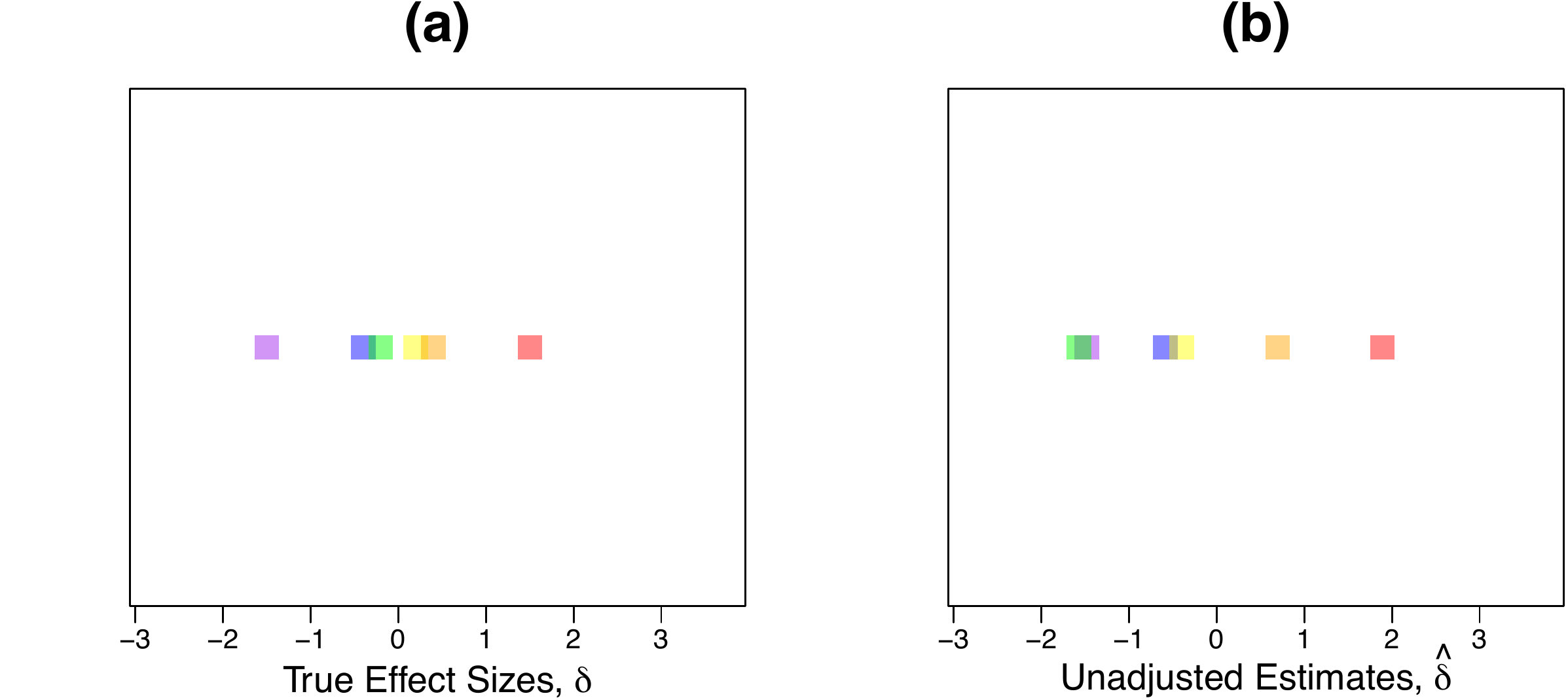}

\vspace{1.5mm}

\includegraphics[scale=0.475]{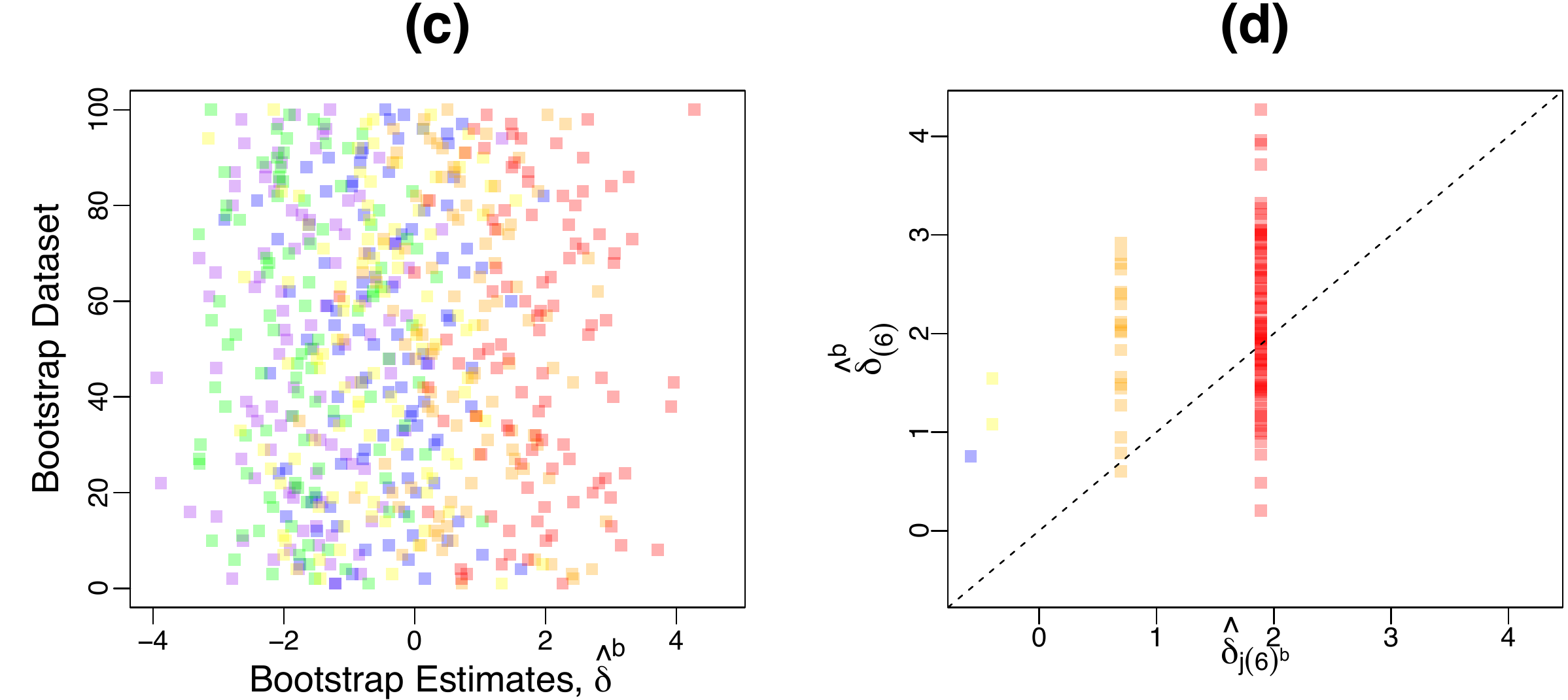}
\vspace{1.5mm}

\includegraphics[scale=0.475]{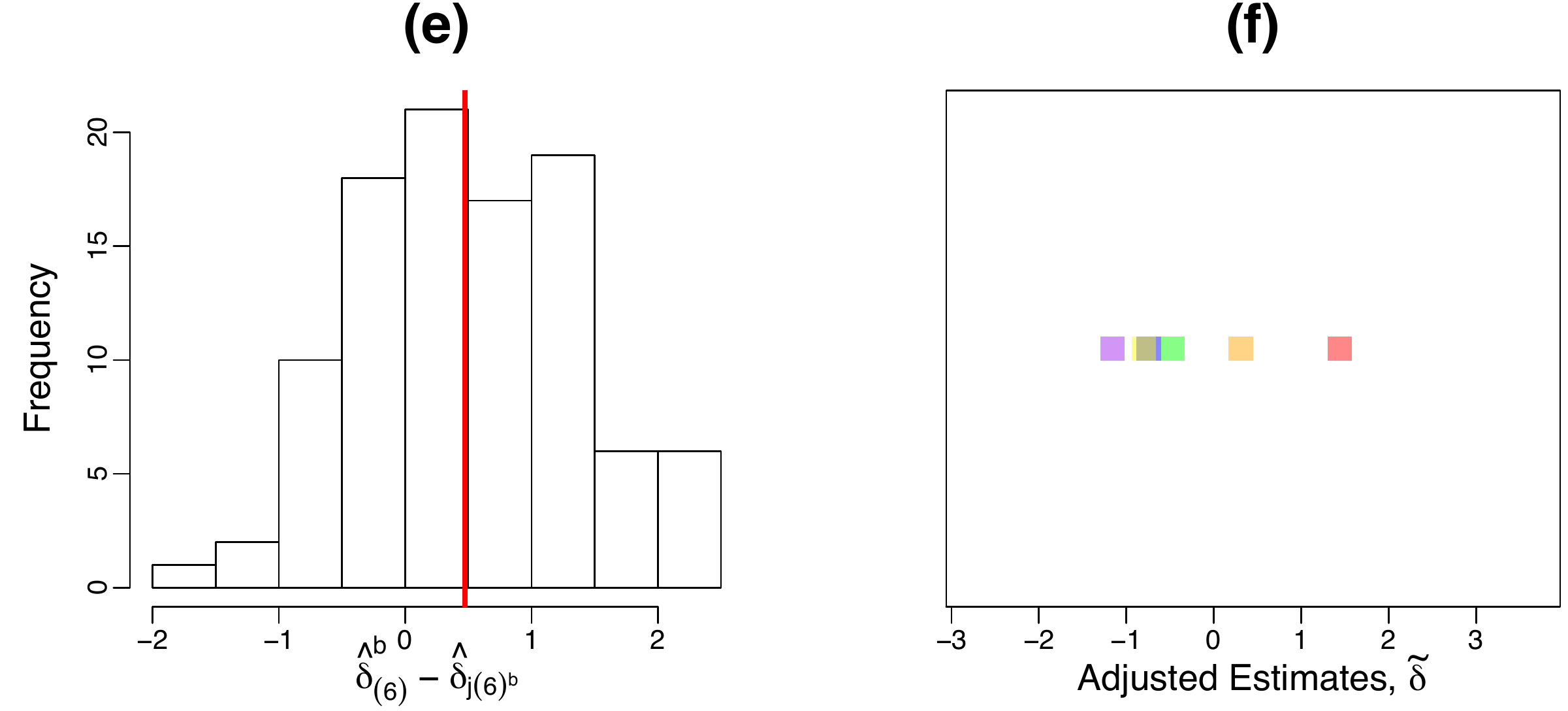}
\end{center}

\caption{\label{Fig:bias plot} \footnotesize  A schematic representation of Algorithm~\ref{Alg:process}: (a) The true effect sizes, $\delta_1,\ldots, \delta_6$. True effect sizes are color-coded with purple and red indicating the smallest and largest effect sizes, respectively.  (b) The unadjusted estimates (Step 1 of Algorithm~\ref{Alg:process}), $\hat{\delta}_1,\ldots,\hat{\delta}_6$.  Unadjusted estimates are color-coded based on their corresponding true effect sizes. (c) 100 bootstrapped effect size estimates simulated using a parametric bootstrap (Step 2 of Algorithm~\ref{Alg:process}).  The 100 bootstrapped effect size estimates  are stacked vertically along the $y$-axis.  (d) Scatterplot of the largest bootstrapped unadjusted estimate ($y$-axis) against the  corresponding unadjusted estimate based on the original data ($x$-axis).  (e) Histogram of the difference between the largest bootstrapped unadjusted estimate and its corresponding unadjusted estimate based on the original data, for 100 bootstrapped data sets (Step 3 of Algorithm~\ref{Alg:process}).  The red vertical line is the mean of the differences, which is the estimate for $\beta_6$. (f) The adjusted effect size estimates $\tilde{\delta}_1,\ldots,\tilde{\delta}_6$ computed using Equation~\ref{Eq:ourestimator}.       }
\end{figure}

\begin{algorithm}[htp]
\caption{Procedure for calculating $\hat{\beta}_k$.}
\begin{enumerate}
\item Calculate the unadjusted estimates $\hat{\boldsymbol \delta} := (\hat{\delta}_1,\ldots,\hat{\delta}_p)^T$.
\item Generate $B$ bootstrapped data sets using the original data set, and obtain unadjusted estimates $\hat{\boldsymbol \delta}^b$ for the $b$th bootstrapped dataset, $b=1,\ldots,B$.
\item Calculate the bias, $\hat{\beta}_k$, as 
the average difference between the $k$th smallest unadjusted estimate of the bootstrapped data, $\hat{\delta}_{(k)}^b$, and the corresponding unadjusted estimate based on the original data, $\hat{\delta}_{j(k)^b}$, where $j(k)^b$ is the index of the $k$th order statistic of $\hat{\boldsymbol{\delta}}^b$.  That is,
\[
\hat{\beta}_k = \frac{1}{B} \sum_{b=1}^B \left( \hat{\delta}_{(k)}^b - \hat{\delta}_{j(k)^b} \right).
\]
\end{enumerate}
\label{Alg:process}
\end{algorithm}
 A schematic representation of the procedure is shown in Figure~\ref{Fig:bias plot}.  Steps 1, 2, and 3 of Algorithm~\ref{Alg:process} are shown in Figures~\ref{Fig:bias plot}(b), \ref{Fig:bias plot}(c), and \ref{Fig:bias plot}(e), respectively.   We propose two bootstrap procedures for Step 2 of Algorithm~\ref{Alg:process}: (1) a parametric bootstrap (Section~\ref{subSec:parametric}), and (2)  a nonparametric bootstrap (Section~\ref{subSec:nonparametric}).
Note that the selection bias problem is apparent in Figure~\ref{Fig:bias plot}.  Figure~\ref{Fig:bias plot}(b) shows that the unadjusted estimates overestimate the largest true effect size and underestimate some of the small true effect sizes near zero.  From Figure~\ref{Fig:bias plot}(f), we see that the adjusted estimates are more accurate than the unadjusted estimates.   

In what follows, we let $\mathbf{D}$ denote the data. 

\subsection{Parametric Bootstrap Approach}
\label{subSec:parametric}
In this section, we present a parametric bootstrap approach for performing Step 2 of Algorithm~\ref{Alg:process}.  This process was described briefly in Section 3 of \citet{noah2013bias}.  This approach assumes the availability of a data-generating model $F(\cdot)$, parametrized by a (possibly multi-dimensional) parameter $\mathbf{\Theta}$. 
The details are in Algorithm~\ref{Alg:parametric}.

\begin{algorithm}[htp]
\caption{Parametric bootstrap for Step 2 of Algorithm~\ref{Alg:process}, {\tt{para}}.}
For $b=1,\ldots,B$:
\begin{enumerate}[(i)]
\item Generate data $\mathbf{D}^b$ according to the data-generating model $F(\hat{\mathbf{\Theta}})$, where $\hat{\mathbf{\Theta}}$ is an estimate of the parameter $\mathbf{\Theta}$.  
\item Calculate the unadjusted estimates $\hat{\boldsymbol{\delta}}^b$ based on $\mathbf{D}^b$.
\end{enumerate}
\label{Alg:parametric}

\end{algorithm}
For concreteness, consider  the case of estimating effect sizes of correlated one-sample $t$-statistics.  The data $\mathbf{D}$ consists of $n$ independent observations and $p$ correlated features.  For instance, if we assume the data-generating model is $F(\cdot) = N(\boldsymbol{\mu},\mathbf{\Sigma})$, then Step (i) of Algorithm~\ref{Alg:parametric} simply amounts to generating each observation of $\mathbf{D}^b$ from a multivariate normal distribution using the sample mean and empirical covariance matrix calculated from $\mathbf{D}$.  

We note that the data-generating model $F(\cdot)$ is not constrained to be a multivariate normal distribution.  For instance, the data might be drawn from a heavy tailed distribution, such as a multivariate $t$-distribution. We rarely know the data-generating model $F(\cdot)$ in practice; this can lead to challenges in applying Algorithm~\ref{Alg:parametric}.  To overcome these challenges, in Section~\ref{subSec:nonparametric} we present an alternative approach.

\subsection{Nonparametric Bootstrap Approach}
\label{subSec:nonparametric}
We now present a nonparametric bootstrap approach for performing Step 2 of Algorithm~\ref{Alg:process}.  Unlike Algorithm~\ref{Alg:parametric}, it does not require knowing the data-generating model.  Since the bootstrapped data is obtained under repeated sampling of the original data, the dependencies among the unadjusted estimates are preserved implicitly.  The details of the proposal are in Algorithm~\ref{Alg:nonparametric}.

\begin{algorithm}[htp]
\caption{Nonparametric bootstrap for Step 2 of Algorithm~\ref{Alg:process}, {\tt{nonpara}}.}
For $b=1,\ldots,B$:
\begin{enumerate}[(i)]
\item Obtain new data  $\mathbf{D}^b$ by resampling independent observations from the data $\mathbf{D}$ with replacement.
\item Calculate the unadjusted estimates $\hat{\boldsymbol{\delta}}^b$ based on $\mathbf{D}^b$.
\end{enumerate}
\label{Alg:nonparametric}
\end{algorithm}

When the data-generating model is complicated or unknown, the nonparametric bootstrap may be more favorable than the parametric approach of Section~\ref{subSec:parametric}.  For instance, in the context of    genome-wide association studies (GWAS) with cases and controls, it is often of interest to estimate the odds ratio for each single nucleotide polymorphism (SNP) within a logistic regression model (see, e.g., \citealp{sun2011br,zhong2010correcting}). 
 Step (i) of Algorithm~\ref{Alg:nonparametric} simply amounts to creating a new data set $\mathbf{D}^b$ by resampling the independent observations from $\mathbf{D}$ with replacement.  Then, a logistic regression is fit, and the odds ratio is computed for each SNP, based on this resampled data.

\section{Properties of the Oracle Estimator}
\label{Sec:Bias Correction Theory}
In this section, we study some properties of the oracle estimator, $\bar{\boldsymbol{\delta}}$,  defined in Equation~\ref{Eq:Frequentist}.  In this section, for simplicity, we assume that the unadjusted estimates $\hat{\boldsymbol{\delta}}$ are unbiased for the true effect sizes $\boldsymbol{\delta}$, and are normally distributed.   
Again, we note that in practice, our proposal does not require normality of the data or of the unadjusted estimates. 

In Section~\ref{Sec:theory1}, we relate the \emph{mean squared error} (MSE) of $\bar{\boldsymbol{\delta}}$ to the MSE of $\hat{\boldsymbol{\delta}}$ and the sum of squared biases of $\hat{\boldsymbol{\delta}}$, and study the effect of correlation among the elements of $\hat{\boldsymbol{\delta}}$ on the sum of squared biases.  In Section~\ref{subSec:importance}, we derive an explicit expression for the amount of reduction in MSE that results from using our proposed estimator as opposed to the estimator of \citet{noah2013bias}, which does not model dependencies in $\hat{\boldsymbol{\delta}}$.

\subsection{Connection between Biases and Correlation}
\label{Sec:theory1}
 We first present a result relating the  MSE of  $\bar{\boldsymbol{\delta}}$ to the MSE of $\hat{\boldsymbol{\delta}}$.

\begin{lemma}
\label{Lemma:noah result} 
\cite{noah2013bias} There is a simple relationship between the sum of squared biases of the unadjusted estimates, the MSE of the oracle estimates, and the MSE of the unadjusted estimates: 
\begin{equation}
\label{Eq:lem1Eq}
\small
\sum_{k=1}^p  E\left[ (\hat{\delta}_k-\delta_{k})^2  \right] = \sum_{k=1}^p \beta_k^2 + \sum_{k=1}^p E\left[  (\bar{\delta}_{k} - \delta_{k})^2 \right].
\end{equation}
\end{lemma}
\noindent In other words, the sum of squared biases $\sum_{k=1}^p \beta_k^2 $ is the amount by which we can improve upon the MSE of $\hat{\boldsymbol{\delta}}$  by correcting for selection bias. 
Note that Lemma~\ref{Lemma:noah result} holds regardless of the correlation among the unadjusted estimates.   

We now study the effect of correlation among the elements of $\hat{\boldsymbol{\delta}}$ on the quantity $\sum_{k=1}^p \beta_k^2$.   We introduce some assumptions and notation that will be used throughout this  section.    We will consider normally distributed unadjusted estimates with mean $\boldsymbol{\delta}$ and various covariance matrices.  To emphasize that the distribution of the unadjusted estimates is a function of their covariance, we will write $\hat{\boldsymbol{\delta}}^{\mathbf{\Sigma}}$ to indicate unadjusted estimates with some arbitrary covariance $\mathbf{\Sigma}$.
Let $\hat{\delta}_{(k)}^{\mathbf{\Sigma}}$ be the $k$th order statistic of the estimates $\hat{\boldsymbol{\delta}}^{\mathbf{\Sigma}}$.  We define $j(k)^{\mathbf{\Sigma}}$ as the index corresponding to the $k$th order statistic of $\hat{\boldsymbol{\delta}}^{\mathbf{\Sigma}}$.  Finally, let $\beta(\boldsymbol{\delta},\mathbf{\Sigma})_k$ be the frequentist selection bias of the $k$th order statistic of $\hat{\boldsymbol{\delta}}^{\mathbf{\Sigma}}$.   
 
The following lemma quantifies the effect of correlation on the bias of the unadjusted estimates.

\begin{lemma}
\label{Lemma:bias}
Let $\small \mathbf{R}$ be an equicorrelation matrix with correlation $\small \rho$ for some $\small -\frac{1}{p-1}< \rho <1$.  Consider two sets of unadjusted estimates, $\small \hat{\boldsymbol{\delta}}^{\mathbf{R}} \sim N(\boldsymbol{\delta},\mathbf{R})$ and $\small \hat{\boldsymbol{\delta}}^{\left({1-\rho} \right)\mathbf{I}} \sim N(\boldsymbol{\delta},\left({1-\rho}\right)\mathbf{I})$.   
For any $\boldsymbol{\delta}$, 
\[
\small \beta(\boldsymbol{\delta},\mathbf{R})_k=\beta(\boldsymbol{\delta},(1-\rho) \mathbf{I})_k.
\]

\end{lemma}

\noindent In other words, the biases in the equicorrelated scenario are equal to the biases in an independent scenario with smaller marginal variance.  
To further explore this, we simplify our model by making the additional assumption that the effect sizes are all equal.

\begin{lemma}
\label{lemma:biggerbias}
Let $\small \hat{\boldsymbol{\delta}}^{\mathbf{R}} \sim N(\boldsymbol{\delta},\mathbf{R})$, where  $\small \mathbf{R}$ is an equicorrelation matrix with correlation $ \small \rho$ for some $ \small   -\frac{1}{p-1}< \rho <1$, and $\small \boldsymbol{\delta} = a\mathbf{1}$ for some constant $\small a$.  Then, 
\begin{equation*}
\begin{split}
\small
 \beta({\boldsymbol{\delta},\mathbf{R}})_k
= \sqrt{1-\rho} \beta(\boldsymbol{\delta},\mathbf{I})_k.
\end{split}
\end{equation*}
\end{lemma}
\noindent The proof is a direct application of Lemma~\ref{Lemma:bias}.  
Lemma~\ref{lemma:biggerbias} implies that as $\rho$ increases, the biases of the unadjusted estimates decrease. Consequently, since by Lemma~\ref{Lemma:noah result}, 
\begin{equation*}
\begin{split}
\sum_{k=1}^p E \left( \bar{\delta}_{k} -\delta_{k} \right)^2  & = \sum_{k=1}^p E\left( \hat{\delta}_k^{\mathbf{R}} - \delta_{k}  \right)^2 - \sum_{k=1}^p \beta(\boldsymbol{\delta},\mathbf{R})^2_k\\
&= \sum_{k=1}^p E\left( \hat{\delta}_k^{\mathbf{R}} - \delta_{k}  \right)^2 - (1-\rho) \sum_{k=1}^p \beta(\boldsymbol{\delta},\mathbf{I})^2_k,
\end{split}
\end{equation*}
and recalling that $\sum_{k=1}^p E(\hat{\delta}_k^{\mathbf{R}} - \delta_k)^2$ is not a function of $\rho$, it follows that the MSE of the oracle estimator increases as $\rho$ increases.

The main implications of our results are as follows. When $\rho$ is approximately zero, existing approaches such as \citet{efron2011tweedie} and \citet{noah2013bias} that do not model correlation in $\hat{\boldsymbol{\delta}}$ can be used to obtain adjusted effect size estimates: in this setting there is no advantage to our proposal.  When $\rho$ is approximately one, the biases of the unadjusted estimates are approximately zero from Lemma~\ref{lemma:biggerbias}, and consequently the MSEs of $\bar{\boldsymbol{\delta}}$ and $\hat{\boldsymbol{\delta}}$ are approximately equal from Lemma~\ref{Lemma:noah result}.  In other words, when $\rho \approx 1$, adjusting the estimates for frequentist selection bias is altogether unnecessary, as the bias is essentially zero.  Our proposal has the potential to gain a substantial amount of reduction in MSE relative to $\hat{\boldsymbol{\delta}}$ and relative to existing approaches that do not model correlation only when there is an intermediate amount of correlation.  We will verify these results empirically in Section~\ref{Sec:Onesample}.

Lemma~\ref{lemma:biggerbias} assumes that $\boldsymbol{\delta}= a \mathbf{1}$.  We now generalize this result by assuming that there are two clusters of effect sizes.   We show that as the separation between the clusters of effect sizes increases, the bias corresponding to each cluster of effect sizes involves only the unadjusted estimates from the corresponding cluster.

\begin{lemma}
\label{theorem:general}
Suppose that $\small \boldsymbol{\delta} = \begin{pmatrix} 
\boldsymbol{\delta}_1\\ \boldsymbol{\delta}_2 \end{pmatrix} =  \begin{pmatrix} \mathbf{0}_{p/2} \\ b \mathbf{1}_{p/2}\end{pmatrix}$ with $\small \hat{\boldsymbol{\delta}}^\mathbf{R}=\begin{pmatrix}   \hat{\boldsymbol{\delta}}_1^\mathbf{R} \\ \hat{\boldsymbol{\delta}}_2^{\mathbf{R}} \end{pmatrix} \sim N \left( \begin{pmatrix} \mathbf{0}_{p/2}\\ b \mathbf{1}_{p/2} \end{pmatrix},  \mathbf{R} \right)$, where $\small \mathbf{R}= \begin{pmatrix} \mathbf{R}_{11} & \mathbf{R}_{12} \\ \mathbf{R}_{21} & \mathbf{R}_{22} \end{pmatrix}$ is any correlation matrix.  Then, as $b \rightarrow \infty$, 
\begin{equation*}
\small
\beta(\boldsymbol{\delta}, \mathbf{R})_k = \begin{cases} \beta(\mathbf{0}_{p/2}, \mathbf{R}_{11})_k & \mathrm{for } \;k = 1,\ldots,\frac{p}{2},\\
 \beta(\mathbf{0}_{p/2}, \mathbf{R}_{22})_k & \mathrm{for } \; k = \frac{p}{2}+1,\ldots,p.
 \end{cases}
\end{equation*}
\end{lemma} 

\noindent A direct consequence of Lemma~\ref{theorem:general} is the following corollary, which generalizes  Lemma~\ref{lemma:biggerbias} to allow for two clusters of effect sizes.

\begin{cor}
\label{cor:two cluster}
Suppose that $\small \hat{\boldsymbol{\delta}}^\mathbf{R}=\begin{pmatrix}   \hat{\boldsymbol{\delta}}_1^{\mathbf{R}} \\ \hat{\boldsymbol{\delta}}_2^{\mathbf{R}} \end{pmatrix} \sim N \left( \begin{pmatrix} \mathbf{0}_{p/2}\\ b \mathbf{1}_{p/2} \end{pmatrix},  \mathbf{R} \right)$,  and $ \small \mathbf{R}=\begin{pmatrix} \mathbf{R}_{11} & \mathbf{R}_{12} \\ \mathbf{R}_{21} & \mathbf{R}_{22}\end{pmatrix}$ is a correlation matrix such that $\small \mathbf{R}_{11} = \mathbf{R}_{22}$ are equicorrelation matrices with correlation $\small \rho$ for some $\small -\frac{1}{p-1} < \rho<1$.  Then, as $\small b\rightarrow \infty$, 
\begin{equation*}
\begin{split}
\small
\beta({\boldsymbol{\delta},\mathbf{R}})_k
= \sqrt{1-\rho} \beta(\mathbf{0}_{p/2},\mathbf{I})_k.
\end{split}
\end{equation*}  
\end{cor}
\noindent We note that both Lemma~\ref{theorem:general} and Corollary~\ref{cor:two cluster} can be generalized to the case of unequal numbers of features in each cluster, more than two clusters of effect sizes, and non-homogenous variances within each cluster.   

\subsection{The Importance of Modelling Dependence}
\label{subSec:importance}
Throughout this section, we will assume that the unadjusted estimates are truly correlated with mean zero, i.e., $\hat{\boldsymbol{\delta}}^{\mathbf{R}}  {\sim} N(\boldsymbol{0},\mathbf{R})$, where $\mathbf{R}$ is an equicorrelation matrix with correlation $\rho$. 
Recall that by definition, $\bar{{\delta}}^{\mathbf{R}}_{j(k)} = \hat{\delta}^{\mathbf{R}}_{(k)} - \beta(\mathbf{0},\mathbf{R})_k $ is the oracle estimator. 
Let
\begin{equation}
\label{Eq:wrong oracle}
\bar{{\delta}}^{\mathbf{I}}_{j(k)} = \hat{\delta}^{\mathbf{R}}_{(k)}-\beta(\mathbf{0},\mathbf{I})_k
\end{equation}
be the ``false oracle" estimator in which the biases $\boldsymbol{\beta}(\mathbf{0},\mathbf{I})$ are computed under the erroneous assumption that the elements of $\hat{\boldsymbol{\delta}}^{\mathbf{R}}$ are uncorrelated. 

Using the results in Section~\ref{Sec:theory1}, we now explore what happens if we erroneously use $\bar{\boldsymbol{\delta}}^{\mathbf{I}}$ as the adjusted estimator instead of $\bar{\boldsymbol{\delta}}^{\mathbf{R}}$. This result is presented in the following lemma. 

\begin{lemma}
\label{lemma:importance}
Suppose that $\small \hat{\boldsymbol{\delta}}^\mathbf{R} \sim N(\mathbf{0},\mathbf{R})$, where $\small \mathbf{R}$ is an equicorrelation matrix with correlation $\small \rho$ for some $\small -\frac{1}{p-1} < \rho < 1$.  Then,
\[
\small
\sum_{k=1}^p E\left[ (\bar{\delta}_{k}^{\mathbf{I}} - \delta_{k})^2\right] =\sum_{k=1}^p E\left[(\bar{\delta}_{k}^{\mathbf{R}} - \delta_{k})^2\right]  + (1-\sqrt{1-\rho})^2 \sum_{k=1}^p  \beta(\mathbf{0},\mathbf{I})_k^2.
\]
\end{lemma} 
\noindent In other words, the oracle estimator $\bar{\boldsymbol{\delta}}^{\mathbf{R}}$, computed under the correct model, dominates the false oracle $\bar{\boldsymbol{\delta}}^{\mathbf{I}}$, computed under the erroneous assumption of no correlation, in terms of MSE.  This motivates our proposal for modeling the correlations among the unadjusted estimates when estimating the biases.

\section{Simulation Studies}
\label{Sec:Onesample}
We consider correcting selection biases of correlated one-sample and two-sample $t$-statistics.  We illustrate \emph{via} simulation studies that failure to account for correlations among the test statistics can give inaccurate effect size estimates. In addition, we consider simulation studies in which the data are generated from a multivariate $t$-distribution.   Furthermore, we show using a two-sample $t$-statistic example that  \verb=nonpara=  is preferable to \verb=para= when the generative model for the data is unknown.

\subsection{Methods and Evaluation of Performance}
\label{subSec:methods}
We compare the following proposals in our simulation studies:
\begin{itemize}
\item \verb=oracle-cor=: oracle estimator assuming that the biases $\beta_k$'s are known (approximated using Monte Carlo), as in Equation~\ref{Eq:Frequentist}.  
\item \verb=oracle-uncor=: the ``false oracle" estimator under the erroneous assumption of no correlation, assuming that the biases $\beta (\mathbf{0},\mathbf{I})_k$'s are known (approximated using Monte Carlo),  as in Equation~\ref{Eq:wrong oracle}.  
\item \verb=tweedie=: empirical Bayes using Lindsey's method, with a spline basis of five degrees of freedom, as the estimate of $\log f(\hat{\delta}_j)$ in Equation~\ref{Eq:posterior} \citep{efron2011tweedie}. 
\item \verb=nlpden=: empirical Bayes using the estimate of $\log f(\hat{\delta}_j)$ in  \citet{wager2013geometric}.
\item \verb=para=: the parametric bootstrap of this paper (Algorithms~\ref{Alg:process} and \ref{Alg:parametric}).
\item \verb=nonpara=: the nonparametric bootstrap approach of this paper (Algorithms~\ref{Alg:process} and \ref{Alg:nonparametric}). 
\item \verb=james-stein=: positive part of the James-Stein estimator,
\[
\hat{\delta}^{\text{JS}}_j = \left(1- \frac{p-2}{\|\hat{\boldsymbol{\delta}} - \bar{\hat{\boldsymbol{\delta}}} \mathbf{1}   \|^2} \right)_+ (\hat{{\delta}}_j-\bar{\hat{\boldsymbol{\delta}}}) + \bar{\hat{\boldsymbol{\delta}}},
\] 
where $(a)_+ = \max(0,a)$, and $\bar{\hat{\boldsymbol{\delta}}}$ is the mean of the unadjusted estimates.
\end{itemize}

\noindent We consider two versions of \verb=para=, which we refer to as \verb=para-uncor= and \verb=para-cor=: in Step (i) of Algorithm~\ref{Alg:parametric},   \verb=para-uncor= generates data assuming that the features are uncorrelated, whereas \verb=para-cor= generates data assuming that the features are correlated (note that \verb=para-uncor= is exactly the proposal of \citet{noah2013bias}).  The specific details for \verb=para-uncor= and \verb=para-cor= are presented in Sections~\ref{subSec:settings} and \ref{subSec:multivariate t} in the context of one-sample $t$-statistics, and in Section~\ref{subSec:twosample} in the context of two-sample $t$-statistics.  Note that out of the methods compared in Sections~\ref{subSec:settings}-\ref{subSec:twosample}, only \verb=para-cor= and \verb=nonpara= account for correlations among the unadjusted test statistics.

In order to evaluate the performances of the methods, we calculate the relative mean squared error (RMSE), defined as the ratio of the MSE of the adjusted effect size estimates to the MSE of the unadjusted estimates: 
\begin{equation}
\label{Eq:MSE}
\text{RMSE} = \frac{\sum (\tilde{\delta}_j - \delta_j)^2}{\sum (\hat{\delta}_j - \delta_j)^2}.
\end{equation}
 The numerator and denominator in Equation~\ref{Eq:MSE} are computed using only the effect sizes for which the unadjusted estimates are most extreme. An RMSE value that is larger than one indicates that the adjusted effect size estimates perform worse than the unadjusted estimates $\hat{\boldsymbol{\delta}}$, and \emph{vice versa}.

\subsection{One-sample $t$-statistics: multivariate Gaussian distribution}
\label{subSec:settings}
Let $\mathbf{D}$ be an $n \times p$ matrix with $n$ independent observations $\mathbf{d}_1,\ldots,\mathbf{d}_n$ and $p$ features.  We define $\hat{\mu}_j $ and $\hat{\sigma}_j$ to be the sample mean and sample standard deviation of the $j$th feature, respectively.  Also, let $\mu_j$ and $\sigma_j$ be the population mean and standard deviation of the $j$th feature, respectively. 
The goal is to estimate the standardized\footnote{In general, the signal-to-noise ratio $\frac{\mu_j}{\sigma_j}$ is of interest.  Here we scale this ratio by $\sqrt{n}$ for convenience. } mean $\delta_j = \sqrt{n}\frac{\mu_j}{\sigma_j}$.  
A natural estimator for the effect size $\delta_j$ is a one-sample $t$-statistic, $\hat{\delta}_j  =\frac{\sqrt{n} \hat{\mu}_j}{\hat{\sigma}_j}$.  It is well known that  $\hat{\delta}_j \stackrel{\cdot}{\sim} N(\delta_j,1)$.  However, if the $p$ features are correlated, then $\hat{\delta}_1,\ldots,\hat{\delta}_p$ will be correlated.

\begin{figure}[htp]
\begin{center}
\includegraphics[scale=0.39]{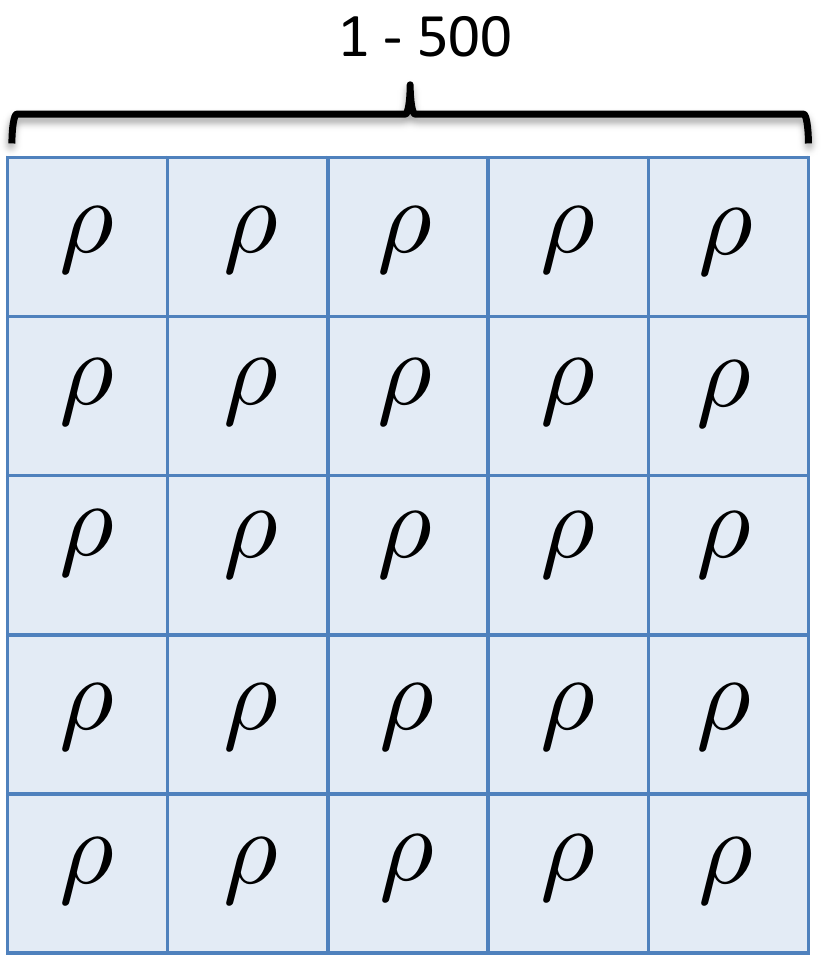}\qquad \qquad 
\includegraphics[scale=0.39]{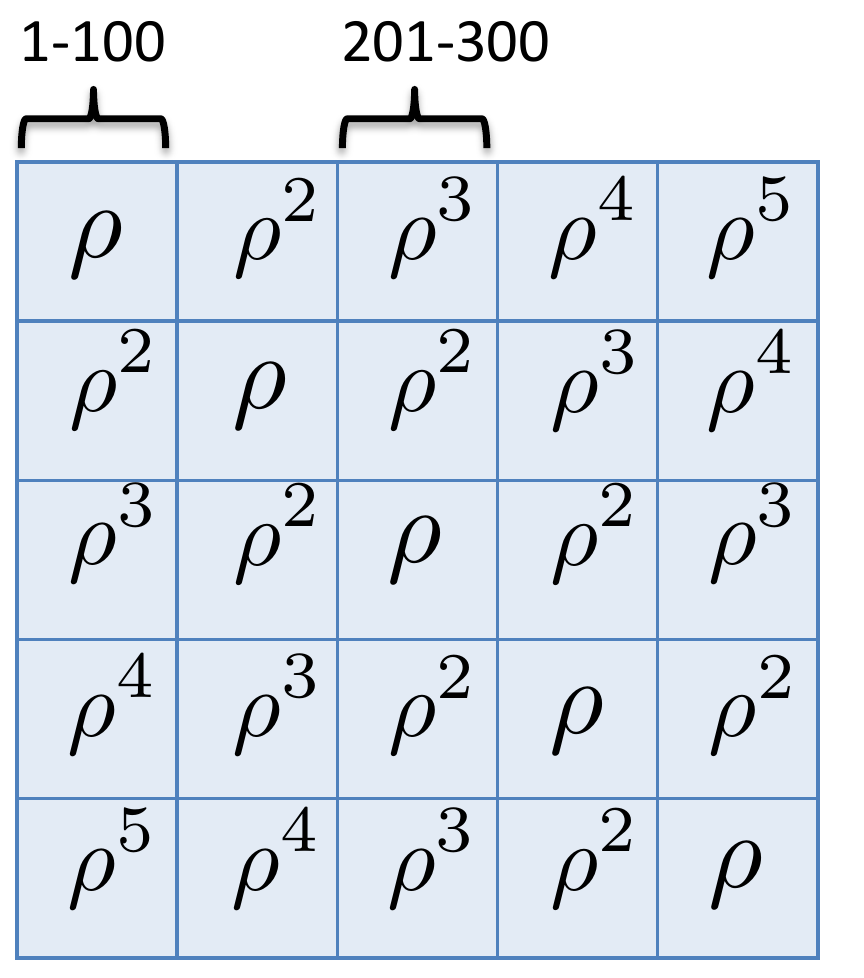}\qquad \qquad 
\includegraphics[scale=0.39]{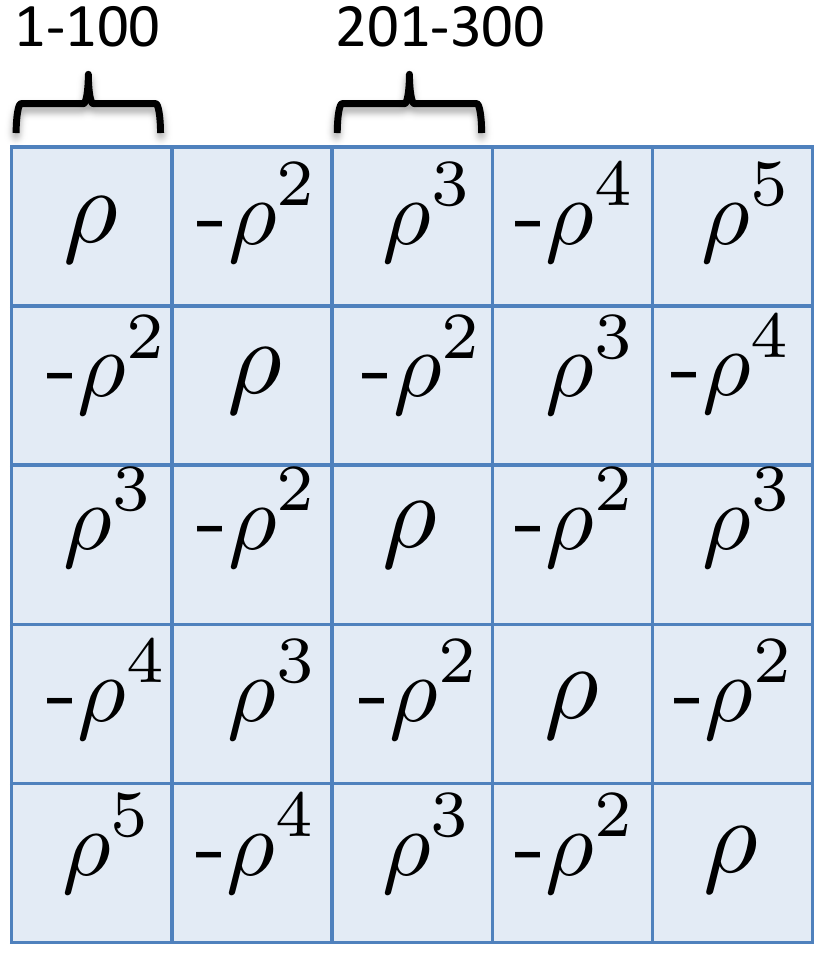}
\end{center}
\caption{\label{Fig:cor structure} Left panel: equicorrelation.  Middle panel:  block AR correlation. Right panel: negative block AR correlation.  For block AR and negative block AR correlation, each block contains 100 features.  The diagonal elements of the correlation matrix are set to equal one.  We consider $\rho \in \{0.5,0.6,0.7,0.8\}$.}
\end{figure}

We generate data according to $\mathbf{d}_1,\ldots,\mathbf{d}_n \stackrel{\mathrm{iid}}{\sim} N(\boldsymbol{\mu}, \mathbf{R})$, where $\mu_j = 0$ for $j=1,\ldots,p-k$, $\mu_j  \stackrel{\mathrm{iid}}{\sim} N(0,0.01)$ for $j=p-k+1,\ldots,p$. 
We consider three different types of correlation structure: (1) equicorrelation, (2) block autoregressive (AR) correlation, and (3) negative block AR correlation, as depicted in Figure~\ref{Fig:cor structure}, with correlation $\rho \in \{0.5,0.6,0.7,0.8\}$.  We use $n=50$, $p=500$, $k=100$, and $\sigma=1$ in our simulation studies.

For \verb=para-uncor= and \verb=para-cor=, we need a generative model in order to apply Step (i) of Algorithm~\ref{Alg:parametric}.  In order to investigate whether ignoring correlations among the test statistics will lead to inaccurate estimates of the effect sizes, for \verb=para-uncor=, we use $D_{ij}^b\stackrel{\text{ind}}{\sim} N(\hat{\mu}_j,\hat{\sigma}_j^2)$, assuming that the features are uncorrelated.  For \verb=para-cor=, we generate the $i$th observation in the $b$th bootstrapped data set from a multivariate normal distribution, $\mathbf{d}_i^b \stackrel{\mathrm{iid}}{\sim} N(\hat{\boldsymbol{\mu}},\hat{\mathbf{\Sigma}})$, where $\hat{\boldsymbol{\mu}}$ and $\hat{\mathbf{\Sigma}}$ are the sample mean and  sample covariance matrix\footnote{In the case of $p>n$, the sample covariance matrix is singular. In order to obtain a positive definite matrix, we add a small constant to the diagonal.}, respectively.  We use $B=1000$ bootstrap samples.

We report the RMSE for the 25 smallest and 25 largest unadjusted test statistics,  averaged over 100 replications, for equicorrelation, block AR, and negative block AR correlation matrices, in Tables~\ref{Table:sim1}-\ref{Table:negblockar},  respectively.

From Table~\ref{Table:sim1}, we see that all approaches perform well when the test statistics are uncorrelated.  We are not paying much of a price even when we allow for correlations using \verb=para-cor= and \verb=nonpara=.  
 When the test statistics are equicorrelated, we see that \verb=para-cor= and \verb=nonpara= outperform other approaches that do not account for correlations among the test statistics.  In addition, the RMSE of \verb=para-cor= is close to the RMSE of \verb=oracle-cor=, which suggests that our estimates of the biases are accurate.  
  We observe that as $\rho$ increases, the RMSEs for \verb=para-cor= and \verb=nonpara= increase from around 0.10 when $\rho=0$ to around 0.55 when $\rho=0.8$.  This is in keeping with our findings in Section~\ref{Sec:Bias Correction Theory} (recall Lemma~\ref{lemma:biggerbias}): the sum of squared biases of the unadjusted estimates decreases as we increase the correlation $\rho$.  Finally, we see that the RMSEs of \verb=oracle-cor= and \verb=oracle-uncor= are the same when $\rho=0$.  As $\rho$ increases, the difference between the RMSE of  \verb=oracle-uncor= and the RMSE of \verb=oracle-cor= increases, with \verb=oracle-uncor= having a larger RMSE.  This agrees with the result presented in Lemma~\ref{lemma:importance}.  
We see similar results from Tables~\ref{Table:blockar} and \ref{Table:negblockar}.  

 \begin{remark}
 One expects \verb=oracle-uncor= to have a lower RMSE than \verb=para-uncor=.  However, we see from Table~\ref{Table:sim1} that when $\rho>0$, this is not the case.  Recall that \verb=para-uncor= is an approximation of \verb=oracle-uncor=, which (by Lemma~\ref{lemma:biggerbias}) overestimates the bias when $\rho >0$.  Also, as pointed out in \citet{noah2013bias}, since the unadjusted estimates are more spread out than the true effect sizes, \verb=para-uncor= ends tends to underestimate the bias relative to \verb=oracle-uncor=.  Therefore, when $\rho>0$, \verb=para-uncor= overestimates the bias less than \verb=oracle-uncor= does, and hence tends to have lower RMSE.  
 \end{remark}

\begin{table}[htp]
\footnotesize
\begin{center}
\caption{The mean RMSE (and standard error) computed as in Equation~\ref{Eq:MSE} using only the 25 smallest and 25 largest unadjusted test statistics, over 100 replications of the simulation study described in Section~\ref{subSec:settings}.  We generate the data with five values of correlation $\rho \in \{ 0,0.5,0.6,0.7,0.8 \}$ in the equicorrelated setting.  For {\tt{para-uncor}}, {\tt{para-cor}}, {\tt{nonpara}}, {\tt{oracle-cor}}, and {\tt{oracle-uncor}}, we use $B=1000$ bootstrap samples. Since $\hat{\delta}_j$ is not an unbiased estimator of $\delta_j$, it may be possible for {\tt{nonpara}} and {\tt{para-cor}} to outperform {\tt{oracle-cor}}; details are in Remark 2.}
\begin{tabular}{| l | c cc cc|}
  \hline
Methods& \multicolumn{5}{c}{Correlation}  \vline   \\ \hline
& 0 &0.5  &0.6  &  0.7 &0.8    \\  \hline
\verb=james-stein= & 0.057 (0.002) &  0.344 (0.013) & 0.464 (0.015) & 0.652 (0.017) & 1.000 (0.026)\\
\verb=tweedie= &0.173 (0.015)&1.036 (0.064)&1.901 (0.128)&4.274 (0.475)&8.761 (1.060)\\
\verb=nlpden=&0.089 (0.004)&0.345 (0.013)&0.462 (0.015)&0.651 (0.017)& 1.002 (0.026)\\
\verb=para-uncor= &0.118 (0.002)&0.341 (0.014)&0.559 (0.019)&1.007 (0.038)& 2.031 (0.101)\\
\verb=para-cor=&0.124 (0.002)&0.299 (0.013)&0.363 (0.014)&0.451 (0.016)&0.575 (0.017)\\
\verb=nonpara=&0.106 (0.002)&0.270 (0.012)&0.334 (0.014)&0.422 (0.016)&0.547 (0.018)\\ \hline
\verb=oracle-cor= &0.057 (0.002)&0.280 (0.014)&0.355 (0.017)&0.455 (0.018)&0.590 (0.019)\\ 
\verb=oracle-uncor= &0.057 (0.002)&0.490 (0.016)&0.770 (0.022)&1.289 (0.043)&2.392 (0.110)\\ \hline
\end{tabular}
\label{Table:sim1}
\end{center}
\end{table}

\begin{table}[htp]
\footnotesize
\begin{center}
\caption{We generate the data with four values of correlation $\rho \in \{ 0.5,0.6,0.7,0.8 \}$ in the block AR setting.  Other details are as in Table~\ref{Table:sim1}. }
\begin{tabular}{| l | c cc c|}
  \hline
Methods& \multicolumn{4}{c}{Correlation}  \vline   \\ \hline
& 0.5 &0.6  &0.7  &  0.8    \\  \hline
\verb=james-stein= & 0.192 (0.010) &  0.259 (0.014) & 0.368 (0.019) & 0.564 (0.028)\\
\verb=tweedie= & 0.558 (0.143) & 0.939 (0.246)&2.089 (0.687)&3.559 (0.439)\\
\verb=nlpden=&0.229 (0.033)&0.274 (0.020)&0.389 (0.023)&0.592 (0.028)\\
\verb=para-uncor= &0.170 (0.009)&0.241 (0.014)&0.405 (0.026)&0.824 (0.059)\\
\verb=para-cor=&0.200 (0.008)&0.243 (0.010)&0.311 (0.013)&0.417 (0.017)\\
\verb=nonpara=&0.177 (0.007)&0.218 (0.009)&0.283 (0.013)&0.387 (0.018)\\ \hline
\verb=oracle-cor= &0.164 (0.009)&0.218 (0.012)&0.299 (0.017)&0.420 (0.022)\\ 
\verb=oracle-uncor= &0.213 (0.013)&0.330 (0.019)&0.555 (0.034)&1.056 (0.067)\\ \hline
\end{tabular}
\label{Table:blockar}
\end{center}
\end{table}

\begin{table}[htp]
\footnotesize
\begin{center}
\caption{We generate the data with four values of correlation $\rho \in \{ 0.5,0.6,0.7,0.8 \}$ in the negative block AR setting.  Other details are as in Table~\ref{Table:sim1}. }
\begin{tabular}{| l | c cc c|}
  \hline
Methods& \multicolumn{4}{c}{Correlation}  \vline   \\ \hline
& 0.5 &0.6  &0.7  &  0.8    \\  \hline
\verb=james-stein= & 0.132 (0.008) &  0.158 (0.010) & 0.196 (0.015) & 0.270 (0.024)\\
\verb=tweedie= & 0.488 (0.148) & 0.690 (0.192)&1.443 (0.434)&2.799 (0.507)\\
\verb=nlpden=&0.148 (0.008)&0.185 (0.013)&0.218 (0.014)&0.303 (0.024)\\
\verb=para-uncor= &0.133 (0.006)&0.161 (0.010)&0.226 (0.022)&0.413 (0.057)\\
\verb=para-cor=&0.156 (0.007)&0.175 (0.009)&0.206 (0.012)&0.267 (0.021)\\
\verb=nonpara=&0.141 (0.006)&0.161 (0.008)&0.192 (0.011)&0.252 (0.020)\\ \hline
\verb=oracle-cor= &0.113 (0.007)&0.140 (0.011)&0.182 (0.017)&0.258 (0.032)\\ 
\verb=oracle-uncor= &0.133 (0.010)&0.186 (0.017)&0.286 (0.032)&0.528 (0.070)\\ \hline
\end{tabular}
\label{Table:negblockar}
\end{center}
\end{table}

 \begin{remark} \label{remark2} 
We expect \verb=oracle-cor= to outperform \verb=para-cor=.  However, in Table~\ref{Table:sim1}, we see that this is not always the case when $\rho$ is large. This is somewhat an artifact of this particular problem. A one-sample $t$-statistic is not an unbiased estimator of the effect size $\delta_j = \sqrt{n} \frac{\mu_j}{\sigma_j}$ (though it is asymptotically unbiased). If we instead use the true value $\sigma_j$ rather than $\hat{\sigma}_j$ in the denominator of our estimator (changing our estimator from one-sample $t$-statistic to one-sample $z$-statistic), then \verb=oracle-cor= would have a lower RMSE than \verb=para-cor=. We provide a more in-depth explanation in the Appendix. 
\end{remark}

\subsection{One-sample $t$-statistics: multivariate $t$-distribution}
\label{subSec:multivariate t}
In this section, we generate data according to a multivariate $t$-distribution, $\mathbf{d}_1,\ldots,\mathbf{d}_n \stackrel{\mathrm{iid}}{\sim} t_\nu(\boldsymbol{\mu},\mathbf{R})$, with location parameter $\boldsymbol{\mu}$,  exchangeable scale matrix  $\mathbf{R}$ with diagonal entries equal to one and off-diagonal entries equal to $\rho$, and degrees of freedom $\nu$.  We note that the covariance matrix of the data, $\mathbf{d}_i$, is $\frac{\nu}{\nu-2} \cdot \mathbf{R}$ \cite{kotz2004multivariate}.
We consider $\nu \in \{10,20\}$ and $\rho \in \{0.6,0.8\}$.  Other simulation details are as in  Section~\ref{subSec:settings}.  Note that \verb=para-uncor= and \verb=para-cor= are performed under the erroneous assumption that the data are multivariate normally distributed, as described in Section~\ref{subSec:settings}.
The RMSE for the 25 smallest and 25 largest unadjusted test statistics, averaged over 100 replications, is in Table~\ref{Table:tdist}.  The results are similar to those of Tables~\ref{Table:sim1}-\ref{Table:negblockar}. 

\begin{table}[htp]
\footnotesize
\begin{center}
\caption{We generate data according to a multivariate $t$-distribution with exchangeable scale matrix, $\mathbf{R}$, with $\rho \in \{0.6,0.8\}$ and degrees of freedom $\nu \in \{10,20\}$.  Other details are as in Table~\ref{Table:sim1}.}
\begin{tabular}{| l | c c|c c|}
  \hline
Methods& \multicolumn{4}{c}{Correlation (degrees of freedom, $\nu$)}  \vline   \\ \hline
& 0.6 (10) &0.8 (10)  &0.6 (20)  &  0.8 (20)    \\  \hline
\verb=james-stein= & 0.427 (0.015) &  0.942 (0.024) & 0.455 (0.016) & 0.977 (0.026)\\
\verb=tweedie= & 1.946 (0.112) & 9.397 (0.714)&1.860 (0.104)&8.349 (0.651)\\
\verb=nlpden=&0.427 (0.015)&0.946 (0.024)&0.456 (0.016)&0.985 (0.026)\\
\verb=para-uncor= &0.552 (0.021)&2.118 (0.102)&0.563 (0.021)&2.061 (0.104)\\
\verb=para-cor=&0.347 (0.014)&0.593 (0.017)&0.362 (0.014)&0.578 (0.017)\\
\verb=nonpara=&0.395 (0.020)&0.643 (0.027)&0.372 (0.017)&0.605 (0.022)\\ \hline
\verb=oracle-cor= &0.343 (0.017)&0.593 (0.020)&0.359 (0.016)&0.598 (0.019)\\ 
\verb=oracle-uncor= &0.751 (0.023)&2.434 (0.108)&0.767 (0.023)&2.399 (0.113)\\ \hline
\end{tabular}
\label{Table:tdist}
\end{center}
\end{table}

\subsection{Two-sample $t$-statistics}
\label{subSec:twosample}
In Section~\ref{subSec:settings}, \verb=nonpara= and \verb=para-cor= performed similarly, since we correctly modeled the data in Step (i) of Algorithm~\ref{Alg:parametric} while performing \verb=para-cor=.  In many scenarios, however, it is unclear how to model the data.  In this section, we consider two versions of \verb=para-cor=:
\begin{itemize}
\item \verb=para-cor-right= in which the data are modeled correctly in Step (i) of Algorithm~\ref{Alg:parametric}.
\item\verb=para-cor-wrong= in which the data are modeled incorrectly in Step (i) of Algorithm~\ref{Alg:parametric}.
\end{itemize}  
We also consider the methods  defined in Section~\ref{subSec:methods}. 
We show empirically that if the data are modeled incorrectly in Step (i) of Algorithm~\ref{Alg:parametric},  even if we allow for dependence among the unadjusted estimates,  \verb=nonpara= may be preferable to \verb=para-cor-wrong=.  We also include results for \verb=para-uncor= to illustrate  that the effect size estimates are inaccurate when we ignore correlations among test statistics.  

Consider an $n_1 \times p$ matrix $\mathbf{D}^{\text{control}}$ and an $n_2 \times p$ matrix $\mathbf{D}^{\text{case}}$ containing independent observations from a control group and a case group, respectively.  Let $\mu_j^{\text{control}}$ and $\mu_j^{\text{case}}$ be the population mean of the $j$th feature for the control group and case group, respectively.  Also, let $\sigma^2_j$ be the common variance of the $j$th feature for the two groups. 

The goal is to estimate the standardized mean difference (up to a scaling factor)   $\delta_j = \frac{\mu_j^{\text{case}}-\mu_j^{\text{control}}}{\sigma_j \sqrt{\frac{1}{n_1}+\frac{1}{n_2}}}$.  A natural estimator for $\delta_j$ is a two-sample $t$-statistic,
\begin{equation}
\label{Eq:twosample}
\hat{\delta}_j = \frac{ \hat{\mu}^{\text{case}}_j- \hat{\mu}^{\text{control}}_j }{\hat{\sigma}_j^{\text{pool}} \cdot \sqrt{\frac{1}{n_1}+\frac{1}{n_2}}},
\end{equation}
where $\hat{\sigma}_j^{\text{pool}} = \sqrt{\frac{(n_1-1)({\hat{\sigma}_j^\text{control}})^2 + (n_2-1)({\hat{\sigma}_j^\text{case}})^2    }{n_1+n_2-2}}$, and $\hat{\sigma}_j^{\text{control}}$ and $\hat{\sigma}_j^{\text{case}}$ are the sample standard deviations of the $j$th feature for the control group and case group, respectively.

We generate data according to $\mathbf{d}_1^{\text{control}}, \ldots, \mathbf{d}_{n_1}^{\text{control}} \stackrel{\text{iid}}{\sim} N(\boldsymbol{\mu}^{\text{control}}, \mathbf{R}^{\text{control}})$ and $\mathbf{d}_1^{\text{case}},\ldots,\mathbf{d}_{n_2}^{\text{case}}  \stackrel{\mathrm{iid}}{\sim} N(\boldsymbol{\mu}^{\text{case}}, \mathbf{R}^{\text{case}})$, where $\mu_j^{\text{control}} \stackrel{\mathrm{iid}}{\sim} N(0,0.01)$ for $j=1,\ldots,p$, $\mu_j^{\text{case}} \stackrel{\mathrm{iid}}{\sim} N(0,0.01)$ for $j=1,\ldots, p-k$ and $\mu_j^{\text{case}} \stackrel{\mathrm{iid}}{\sim} N(0.5,0.01)$ for $j=p-k+1,\ldots,p$. We let $\mathbf{R}^{\text{control}}$ be an equicorrelation matrix with correlation 0.5.  In addition, we construct $\mathbf{R}^{\text{case}}$ as 
\begin{equation*}
R_{jj'}^{\text{case}} = \begin{cases} 1 & \text{if } j=j',\\
0.8 & \text{if } j\ne j' \text{ and } j,j'\in\left\{1\ldots,p-k\right\},\\
0.8 & \text{if } j\ne j' \text{ and } j,j'\in \{p-k+1,\ldots,p\},\\
0.5 & \text{otherwise}.
\end{cases}
\end{equation*}
We consider $n_1 =40$, $n_2=40$, $p=500$, $k=200$, and $B=1000$.

For \verb=para-uncor=, we independently generate each element of $\mathbf{D}^b$ in Step (i) of Algorithm~\ref{Alg:parametric} according to a normal distribution with  sample mean and sample standard deviation corresponding to the group to which the observation belongs.  For \verb=para-cor-right=, in Step (i) of Algorithm~\ref{Alg:parametric}, we generate each observation according to a multivariate normal distribution with sample mean and sample covariance corresponding to the group to which the observation belongs.
For \verb=para-cor-wrong=, we intentionally misspecify the data-generating model.  We assume that features from both groups share the same covariance matrix, estimated using a pooled covariance matrix.  Therefore, for \verb=para-cor-wrong=, in Step (i) of Algorithm~\ref{Alg:parametric}, each observation is generated according to a multivariate normal distribution with sample mean corresponding to the group to which the observation belongs,  and a pooled sample covariance matrix.  
We report the RMSE for the 25 smallest and 25 largest unadjusted test statistics, averaged over 100 replications, in Table~\ref{Table:twosample}.  

\begin{table}[htp]
\small
\begin{center}
\caption{The mean RMSE (and standard error) for the 25 smallest and 25 largest unadjusted test statistics, over 100 replications of the simulation study described in Section~\ref{subSec:twosample}. Here, {\tt{para-cor-right}} and {\tt{para-cor-wrong}} were performed using the correct and wrong model in Step (i) of Algorithm~\ref{Alg:parametric}, respectively.  For {\tt{para-uncor}}, {\tt{para-cor-right}}, {\tt{para-cor-wrong}}, {\tt{nonpara}}, {\tt{oracle-cor}}, and {\tt{oracle-uncor}}, we use $B=1000$ bootstrap samples.}
\begin{tabular}{| l | c|}
  \hline
Methods &  RMSE   \\ \hline
\verb=james-stein= &1.350 (0.138) \\
\verb=tweedie= & 2.283 (0.206) \\
\verb=nlpden= &  1.262 (0.109)\\
\verb=para-uncor=  & 1.188 (0.098)\\
\verb=para-cor-wrong=  &0.662 (0.028)\\
\verb=para-cor-right=  &0.538 (0.028) \\
\verb=nonpara=  &  0.562 (0.028)\\ \hline
\verb=oracle-cor= &0.580 (0.031)\\
\verb=oracle-uncor= &  1.383 (0.104)\\
   \hline
   
\end{tabular}
\label{Table:twosample}
\end{center}
\end{table}

 We see from Table~\ref{Table:twosample} that \verb=nonpara= and \verb=para-cor-right= perform  significantly better than \verb=para-cor-wrong=. This is because the pooled empirical covariance matrix used by \verb=para-cor-wrong= is not an accurate estimate of the correlations among features in either group.  This example may seem somewhat contrived --- since we simulated the data, we know that the correlations are different in each group, and we could have gotten better results using \verb=para-cor-right= by making use of this information.  However, in practice, we rarely know the true underlying model for real data.

We have shown that for \verb=para-cor=, the estimated effect sizes are very sensitive to model misspecification in Step (i) of Algorithm~\ref{Alg:parametric}.  The \verb=nonpara= approach is more appealing, since it implicitly models the correlations among features in each group.

\section{Application to Gene Expression Data}
\label{Sec:Realdata}
We now consider two gene expression data sets: 
\begin{enumerate}
\item Prostate data \citep{singh2002gene}.  The data set contains gene expression levels of $p=6033$ genes  on 50 controls and 52 males with prostate cancer.
\item Lung cancer data \citep{spira2007airway}.  The data set contains gene expression levels from large airway epithelial cells on 90 controls and 97 patients with lung cancer.  The data set can be found in Gene Expression Omnibus at accession number GDS2771 \citep{barrett2007ncbi}.   There are $p=22283$ genes in this data set.  We consider only the 5000 genes with largest standard deviation.  
\end{enumerate}
We apply the methods described in Section~\ref{subSec:methods} to obtain bias-corrected two-sample $t$-statistics.  For \verb=para-uncor= and \verb=para-cor=, we generate data $\mathbf{D}^b$ in Step (i) of Algorithm~\ref{Alg:parametric} according to a multivariate normal distribution with the sample mean and the sample covariance matrix corresponding to each group; for \verb=para-uncor=, we assume that the sample covariance matrix is diagonal.

In order to compare the performances of the different methods, we randomly split the data into equal-sized training and test sets.  We then calculate the sum of squared differences between the estimated effect sizes from the training set and the unadjusted estimates from the test set, 
\begin{equation}
\label{Eq:testMSE}
\sum (\tilde{\delta}_j^{\text{train}}-\hat{\delta}_j^{\text{test}})^2,
\end{equation}
where $\tilde{\delta}_j^{\text{train}}$ is the $j$th adjusted estimate from the training set and $\hat{\delta}_j^{\text{test}}$ is the $j$th unadjusted estimate from the test set.  In Equation~\ref{Eq:testMSE}, the summation is taken over the features corresponding to the $k$ smallest and $k$ largest unadjusted estimates on the training set.  We consider $k=\{15,25,50\}$. Small values of this quantity indicate that the bias-corrected estimates are close to the true effect sizes.  A similar approach is taken in \citet{ferguson2013empirical}.  The results for the prostate data and the lung cancer data, averaged over 100 random splits of the data,  are reported in Table~\ref{Table:prostatedata} and Table~\ref{Table:spiradata}, respectively.

\begin{table}[htp]
\small
\begin{center}
\caption{Results for prostate data: the mean sum of squared differences (and standard error) between the estimated effect sizes  from the training set and the unadjusted estimates from the test set as in Equation~\ref{Eq:testMSE}, for features  corresponding to the $k$ smallest and $k$ largest unadjusted estimates on the training set, over 100 random splits of the data.  We use $B= 1000$ bootstrap samples for {\tt{para-uncor}}, {\tt{para-cor}}, and {\tt{nonpara}}.}
\begin{tabular}{| l | c|c | c| }
  \hline
Methods & $k=50$ & $k=25$  &  $k=15$  \\ \hline
\verb=james-stein= & 190.92 (2.34) & 97.60 (1.69)  & 58.06 (1.26) \\
\verb=nlpden= & 216.19 (3.67) & 119.47 (3.08) &78.01 (2.66)\\
\verb=tweedie= & 204.33 (3.11)&110.13 (2.83) &71.03 (2.40)\\
\verb=para-uncor=  & 190.81 (2.40)& 93.56 (1.84)&54.93 (1.40)\\
\verb=para-cor=  &178.65 (1.97)&87.90 (1.55)&51.07 (1.17)\\
\verb=nonpara=  &191.73 (2.42)&93.65 (1.84)&54.75 (1.37)\\ \hline
unadjusted estimates  & 729.62 (8.05) &400.35 (5.76) & 258.56 (4.21) \\ \hline
   
\end{tabular}
\label{Table:prostatedata}
\end{center}
\end{table}

\begin{table}[htp]
\small
\begin{center}
\caption{Results for lung cancer data.  Details are as in Table~\ref{Table:prostatedata}.}
\begin{tabular}{| l | c|c | c| }
  \hline
Methods & $k=50$ & $k=25$&  $k=15$ \\ \hline
\verb=james-stein= & 267.91 (19.56) & 137.63 (9.95) & 82.25 (6.06) \\
\verb=nlpden= & 230.88 (17.83) & 115.24 (8.65) &68.13 (5.09)\\
\verb=tweedie= & 254.68 (18.07)&128.92 (9.02) &76.00 (5.43)\\
\verb=para-uncor= & 209.18 (13.76)& 105.85 (6.71)&63.20 (3.97)\\
\verb=para-cor=  &206.41 (15.88)  &  103.00 (7.91)& 60.96 (4.63)\\
\verb=nonpara=  &201.59 (15.04)&100.52 (7.45)&59.38 (4.35)\\ \hline
unadjusted estimates &388.76 (29.00) &215.00 (15.86) &138.83 (9.83) \\ \hline
\end{tabular}
\label{Table:spiradata}
\end{center}
\end{table}

From Table~\ref{Table:prostatedata}, we see that \verb=nlpden= and \verb=tweedie= have substantially worse performance than the other methods.  This is because both \verb=nlpden= and \verb=tweedie= involve obtaining a smooth estimate of the marginal density: there simply are not enough extreme unadjusted estimates to obtain an accurate estimate of the marginal density.  Surprisingly, \verb=para-uncor=, \verb=nonpara=, and \verb=james-stein= have similar performance.  We believe that the poor performance of \verb=nonpara= relative to \verb=para-cor= is due to the small sample size after splitting the data set into training and test sets.  Finally, we observe that \verb=para-cor= has the lowest sum of squared differences.   This is because the normality assumption of the data-generating model in Step (i) of Algorithm~\ref{Alg:parametric} is approximately valid, since this microarray data set is pre-processed and normalized.    This implies that accounting for correlations leads to more accurate effect size estimates.  
We see similar results for the lung cancer data set in Table~\ref{Table:spiradata}.

\section{Discussion}
\label{Sec:Discussion}
In this paper, we have extended the framework of \citet{noah2013bias} in order to correct  dependent unadjusted estimates for selection bias.  We proposed a nonparametric bootstrap and a parametric bootstrap procedure for this purpose.  Unlike existing proposals, our proposal largely avoids the need for parametric assumptions about the unadjusted estimates.  Therefore, it is applicable to quite general scenarios in which the data or the unadjusted estimates are not normally distributed.

  An interesting question for future work is whether the results in Section~\ref{Sec:Bias Correction Theory} can be generalized to the case of an  arbitrary mean vector and  covariance matrix, under some mild assumptions.  Additionally, one might explore a potential connection between our proposed framework for bias correction, and the false discovery rate. \citet{efron2011tweedie} discussed the connection between the empirical Bayes method and the false discovery rate.

\section*{Acknowledgment}
We thank Stefan Wager for providing us \verb=R=-code for implementing \verb=nlpden=. D.W. was partially supported by a Sloan Research Fellowship, NIH Grant DP5OD009145, and NSF CAREER DMS-1252624.  N.S. was partially supported by NIH Grant DP5OD019820.  The content is solely the responsibility of the authors and does not necessarily represent the official views of the funding agencies.

\bibliography{reference}

\appendix
\section*{Appendix}
This section provides a more extensive discussion on Remark~\ref{remark2} in the manuscript and the proofs of the results from Section~\ref{Sec:Bias Correction Theory}.

\subsection*{A More Extensive Discussion on Remark~\ref{remark2}:}
Intuitively, we expect \verb=oracle-cor= to have a lower RMSE than \verb=para-cor=.  From Table~\ref{Table:sim1}, we see that this is not the case when $\rho$ is large.  We provide a heuristic argument for this observation.  
For simplicity, consider the case when $\rho\approx 1$ and $\boldsymbol{\delta} = \mathbf{0}$.  Suppose that $\hat{\delta}_j >0$.  Then, using Jensen's Inequality and the fact that $(\hat{\sigma}_j^b)^2| \hat{\sigma}_j \sim  \hat{\sigma}_j^2 \chi^2_1$, we obtain
\begin{equation}
\label{Eq:oracle bad behavior}
\small
E\left[\hat{\delta}_j^b\Big{|}\bar{x}_j,\hat{\sigma}_j\right] - \hat{\delta}_j= \sqrt{n} \bar{x}_j E \left[\frac{1}{\hat{\sigma}_j^b} \Big{|}\hat{\sigma}_j\right] -\hat{\delta}_j \ge\sqrt{n} \bar{x}_j  \frac{1}{E(\hat{\sigma}_j^b|\hat{\sigma}_j)} - \hat{\delta}_j  =  \sqrt{\frac{\pi}{2}} \hat{\delta}_j -\hat{\delta}_j   >     0.
\end{equation}
By definition, the bias estimate for the $k$th order statistic of \verb=para-cor= is $\hat{\beta}_k =\frac{1}{B}\sum_{b=1}^B (\hat{\delta}_{(k)}^b - \hat{\delta}_{j(k)^b})$.  Since $\rho \approx 1$, we know that $\hat{\delta}_j \approx \hat{\delta}_{j'}\; \forall j,j' \in \{1,\ldots, p\}$ and hence, $\hat{\beta}_k$ is a good approximation to the left-hand-side of (\ref{Eq:oracle bad behavior}).  Therefore, the bias estimates  from \verb=para-cor= tend to have the same signs as the unadjusted estimates and tend to be positively correlated with the unadjusted estimates.  In contrast, by definition, biases from  \verb=oracle-cor=  are uncorrelated with  the unadjusted estimates.

Now, since $\boldsymbol{\delta}=0$, the numerator of the RMSE in (\ref{Eq:MSE}) for the \verb=para-cor= adjusted estimates is $\sum_{j=1}^p \tilde{\delta}_j^2 = \sum_{k=1}^p (\hat{\delta}_{(k)} - \hat{\beta}_k )^2$.  In contrast, the numerator of the RMSE for \verb=oracle-cor= is $\sum_{k=1}^p  (\hat{\delta}_{(k)}-\beta_k)^2$.  Since $\beta_k$ and $\hat{\delta}_{(k)}$ for \verb=oracle-cor= are uncorrelated, and $\hat{\beta}_k$ and $\hat{\delta}_{(k)}$ for \verb=para-cor= are positively correlated,  the RMSE of  \verb=para-cor= is lower than the RMSE of \verb=oracle-cor= when $\rho \approx 1$.  

The inequality in (\ref{Eq:oracle bad behavior}) is due to the fact that a one-sample $t$-statistic is not an unbiased estimate of the effect size.  If we were to use a one-sample $z$-statistic, then \verb=oracle-cor= would have a lower RMSE than \verb=para-cor=, even when $\rho\approx 1$.

\subsection*{Proof of Lemma~\ref{Lemma:bias}:}

\begin{proof}
Let $\epsilon_0 \sim N(0,1)$, $\epsilon_j \stackrel{\mathrm{iid}}{\sim} N(0,1)$, and  $E[\epsilon_0 \epsilon_j] = \lambda$ for some constant $\lambda$.  From \citet{owen1962moments}, the model $\hat{\boldsymbol{\delta}}^{\mathbf{R}} \sim N(\boldsymbol{\delta},\mathbf{R})$ can be rewritten as 
\begin{equation}
\label{Eq:corbias}
\hat{\delta}_j^{\mathbf{R}} = \delta_j +a \epsilon_0 + \left(\sqrt{1-\rho} \right)\epsilon_j, \qquad \text{ for } j=1,\ldots, p.
\end{equation}
for some constant $a$.
\citet{owen1962moments} showed that when $\rho\ge 0$, the constants take values $\lambda = 0$ and $a=\sqrt{\rho}$.  In contrast, when $\rho = -\alpha^2 <0$ for some $\alpha <\sqrt{\frac{1}{p-1}}$, the constants take values $\lambda = -\frac{\alpha}{\sqrt{1-\rho}}$ and $a=\alpha$. 

 Next, we note that the model $\hat{\boldsymbol{\delta}}^{(1-\rho)\mathbf{I}} \sim N(\boldsymbol{\delta},\left({1-\rho}\right)\mathbf{I})$ can be rewritten as 
\begin{equation}
\label{Eq:identitybias}
\hat{\delta}_j^{\left( 1-\rho \right)\mathbf{I}} = \delta_j  +\left( \sqrt{1-\rho}\right) \epsilon_j \qquad \text{ for } j=1,\ldots, p.
\end{equation}

We now prove the assertion $\beta(\boldsymbol{\delta},\mathbf{R})_k=\beta(\boldsymbol{\delta},(1-\rho) \mathbf{I})_k$.  By the definition of frequentist selection bias, we have 
\begin{equation*}
\begin{split}
\beta(\boldsymbol{\delta},\mathbf{R})_k &= E[\hat{\delta}_{(k)}^{\mathbf{R}}-\delta_{j(k)^{\mathbf{R}}}]\\
&= E\left[\delta_{j(k)^{\mathbf{R}}} + a\epsilon_0 + \left(\sqrt{1-\rho}\right) \epsilon_{j(k)^{\mathbf{R}}}-\delta_{j(k)^{\mathbf{R}}}\right]\\
&=\left(\sqrt{1-\rho}\right) E[\epsilon_{j(k)^\mathbf{R}}].
\end{split}
\end{equation*}
Similarly, 
\begin{equation*}
\begin{split}
\beta(\boldsymbol{\delta},\left( 1-\rho \right)\mathbf{I})_k &= E\left[\hat{\delta}_{(k)}^{\left( 1-\rho \right)\mathbf{I}}-\delta_{j(k)^{\left( 1-\rho \right)\mathbf{I}}}\right]\\
&= E\left[\delta_{j(k)^{\left( 1-\rho \right)\mathbf{I}}}  + \left(\sqrt{1-\rho}\right) \epsilon_{j(k)^{\left( 1-\rho \right)\mathbf{I}}}-\delta_{j(k)^{\left( 1-\rho \right)\mathbf{I}}}\right]\\
&=\left(\sqrt{1-\rho}\right) E\left[\epsilon_{j(k)^{\left( 1-\rho \right)\mathbf{I}}}\right].
\end{split}
\end{equation*}
Since (\ref{Eq:corbias}) and~(\ref{Eq:identitybias}) indicate that $\hat{\delta}_j^{\mathbf{R}}$ and $\hat{\delta}_j^{(1-\rho)\mathbf{I}}$ differ only by a location shift, and a location shift does not change the indices of the order statistics, $j(k)^{\left( 1-\rho \right)\mathbf{I}}$ has the same distribution as $j(k)^{\mathbf{R}}$.  Therefore, the expectation of $\epsilon_{j(k)^{\mathbf{R}}}$ is the same as the expectation of $\epsilon_{j(k)^{\left( 1-\rho \right) \mathbf{I}}}$.
\end{proof}

\subsection*{Proof of Lemma~\ref{theorem:general}:}

\begin{proof}
Without loss of generality, we prove the result for $k=1,\ldots,\frac{p}{2}$.  We note that the results for $k=\frac{p}{2}+1,\ldots,p$ can be obtained by a similar argument.  Let 
\[
A = \left[ \left\{\max(\hat{\boldsymbol{\delta}}_1) \le \frac{b}{2} \right\}  \cap \left\{\min(\hat{\boldsymbol{\delta}}_2) \ge \frac{b}{2}   \right\} \right].
\]  
Then, by symmetry, union bound, and Gaussian tail inequality,
\begin{equation}
\label{Eq:pac3}
\begin{split}
P(A^c) &\le  P\left( \max ( \hat{\boldsymbol\delta}_1 ) > \frac{b}{2} \right) + P\left( \min ( \hat{\boldsymbol\delta}_2 ) < \frac{b}{2} \right)\\
&\le p \cdot P\left(N(0,1) > \frac{b}{2}\right)\\
&\le  e^{-\frac{b^2}{8} - \log b + \log p}.
\end{split}
\end{equation}

 We now calculate the frequentist selection bias for the $k$th order statistic
\begin{equation*}
\begin{split}
\beta(\boldsymbol{\delta},\mathbf{I})_k &= E\left[\hat{\delta}_{(k)} - \delta_{j(k)}\right]\\
&=E\left[\left( \hat{\delta}_{(k)} - \delta_{j(k)} \right) 1_{A}\right] + E\left[\left( \hat{\delta}_{(k)}  -\delta_{j(k)} \right)1_{A^c}\right] \\
&= I + II.
\end{split} 
\end{equation*}

\noindent The goal is to show that $I$ converges to $\beta(\boldsymbol{\delta}_1,\mathbf{R}_{11})_k $ for $k=1,\ldots, \frac{p}{2}$,  and $II$ converges to zero as $b\rightarrow \infty$.  

We define some additional notation that will be used throughout the proof.  Let $\hat{\delta}_{1,(k)}$ and $\delta_{1,j(k)}$ be the $k$th order statistic of $\hat{\boldsymbol{\delta}}_1$ and its  corresponding effect size, respectively.  

$I$: For $k=1,\ldots,\frac{p}{2}$, note that  $\delta_{j(k)} 1_{A} =0$ and $\hat{\delta}_{(k)} 1_{A} = \hat{\delta}_{1,(k)}1_{A}$.  Also, note that $|\hat{\delta}_{1,(k)} 1_{A}| \le |\hat{\delta}_{1,(k)}|$ and that $E[|\hat{\delta}_{1,(k)}|] < \infty$.  Thus, by the Dominated Convergence Theorem,
\begin{equation*}
\begin{split}
E\left[\left( \hat{\delta}_{(k)} -\delta_{j(k)} \right) 1_A\right]   &= E\left[\hat{\delta}_{1,(k)}  1_{A}\right] \\
&\rightarrow E [\hat{\delta}_{1,(k)} ] \\
&= \beta(\boldsymbol{\delta}_1,\mathbf{R}_{11})_k.
\end{split}
\end{equation*}

$II$: We will now show that $\left|E\left[ \left( \hat{\delta}_{(k)} -\delta_{j(k)}  \right) 1_{A^c}\right]\right| \rightarrow 0$, implying $II\rightarrow 0$.  The fact that $
|\hat{\delta}_{(k)}| \le \max \left\{ |\hat{\delta}_{1,(\frac{p}{2})}|, |\hat{\delta}_{2,(\frac{p}{2})}|    \right\} \le |\hat{\delta}_{1,(\frac{p}{2})}| + |\hat{\delta}_{2,(\frac{p}{2})}|$
implies that
\begin{equation}
\label{Eq:fact}
 |\hat{\delta}_{(k)} - \delta_{j(k)}| \le |\hat{\delta}_{1,(\frac{p}{2})}| + |\hat{\delta}_{2,(\frac{p}{2})}|  + |b| \le |\hat{\delta}_{1,(\frac{p}{2})}| + |\hat{\delta}_{2,(\frac{p}{2})}-b|  + 2|b|.
 \end{equation}

\noindent Also, note that the event 
\begin{equation*}
\begin{split}
A^c &= \left[ \left\{\max(\hat{\boldsymbol{\delta}}_1) > \frac{b}{2} \right\}  \cup \left\{\min(\hat{\boldsymbol{\delta}}_2) < \frac{b}{2}   \right\} \right]\\
&=  \left[ \left\{\max(\hat{\boldsymbol{\delta}}_1) > \frac{b}{2} \right\}  \cup \left\{\min(\hat{\boldsymbol{\delta}}_2 -b\mathbf{1}_{p/2} ) < -\frac{b}{2}   \right\} \right],\\
\end{split}
\end{equation*}
and that
\[
\left||\hat{\delta}_{1,(\frac{p}{2})}  |1_{A^c}\right| \le |\hat{\delta}_{1,(\frac{p}{2})}  | \qquad \text{and} \qquad E\left[ |\hat{\delta}_{1,(\frac{p}{2})}  | \right] < \infty,
\]
\[
\left| |\hat{\delta}_{2,(\frac{p}{2})} -b | 1_{A^c}\right| \le  |\hat{\delta}_{2,(\frac{p}{2})}-b | \qquad \text{and} \qquad E\left[ |\hat{\delta}_{2,(\frac{p}{2})}-b|\right] < \infty.
\]

 By Jensen's Inequality and the above facts, 
\begin{equation*}
\begin{split}
\left|E[\left(\hat{\delta}_{(k)}  -\delta_{j(k)}\right) 1_{A^c}]\right|  &\le E\left[|  \hat{\delta}_{(k)} -\delta_{j(k)} | 1_{A^c}\right]\\
&\le E\left[ |\hat{\delta}_{1,(\frac{p}{2})}|  1_{A^c}\right]+ E\left[|\hat{\delta}_{2,(\frac{p}{2})} - b|  1_{A^c}\right] + 2|b| P(A^c)\\
 &\rightarrow 0,
\end{split}
\end{equation*}
where the limit statement is obtained by (\ref{Eq:pac3}) and the Dominated Convergence Theorem.

\end{proof}
\subsection*{Proof of Lemma~\ref{lemma:importance}:}
\begin{proof}
By Lemma~\ref{Lemma:bias}, we have that $\beta(\mathbf{0},\mathbf{R})_k = \sqrt{1-\rho} \cdot \beta(\mathbf{0},\mathbf{I})_k$.  Thus, the mean squared error for the ``false oracle" estimator $\bar{\boldsymbol{\delta}}^{\mathbf{I}}$ is 
\begin{equation*}
\begin{split}
\sum_{k=1}^p E\left[(\bar{\delta}_{j(k)}^{\mathbf{I}} - \delta_{j(k)})^2\right]  &= \sum_{k=1}^p E\left[(\hat{\delta}_{(k)}^{\mathbf{R}} - \beta(\mathbf{0},\mathbf{I})_k- \delta_{j(k)} -\beta(\mathbf{0},\mathbf{R})_k +\beta(\mathbf{0},\mathbf{R} )_k)^2\right]  \\
&= \sum_{k=1}^p E\left[(\bar{\delta}_{j(k)}^{\mathbf{R}} - \delta_{j(k)}  - [1-\sqrt{1-\rho}]  \cdot \beta(\mathbf{0},\mathbf{I})_k)^2\right]\\
&= \sum_{k=1}^p E\left[(\bar{\delta}_{j(k)}^{\mathbf{R}} - \delta_{j(k)})^2 \right]  + (1-\sqrt{1-\rho})^2 \sum_{k=1}^p \beta(\mathbf{0},\mathbf{I})_k^2,
\end{split}
\end{equation*}
since the expected value of the cross-term equals zero.
\end{proof}


\end{document}